\documentstyle[11pt,aaspp4,psfig,epsfig]{article}
\begin{document}

%\slugcomment{Resubmitted to ApJ}
%\begin{flushleft}
%Revised Manuscript MS\#57623
%\end{flushleft}

\lefthead{You et al.}
\righthead{ Does the Iron K$_{\alpha}$ Line of Active Galactic Nuclei 
Arise from the Cerenkov Line-like Radiation? }

\title{
Does the Iron K$_{\alpha}$ Line of Active Galactic Nuclei  
Arise from the Cerenkov Line-like Radiation? }

\author{\small \it J. H. You\altaffilmark{1, 2}, D. B. Liu\altaffilmark{1},
W. P.~ Chen\altaffilmark{3}, L. Chen\altaffilmark{1}, 
S. N. Zhang\altaffilmark{4,5,6}
\affil{$^{1}$Institute for Space and Astrophysics, Department of
Physics, Shanghai Jiao-Tong University, Shanghai, 200030, China, P. R.}
\affil{$^{2}$jhyou@online.sh.cn}
\affil{$^{3}$Institute of Astronomy and Department of Physics, 
National Central University, Chung-Li, 32054, Taiwan, China}
\affil{$^{4}$Center for Astrophysics, Physics Department, Tsinghua University,
 Beijing, 100084, China, P. R.}
\affil{$^{5}$Physics Department, University of Alabama in Huntsville, 
Huntsville, AL 35899, USA}
\affil{$^{6}$National Space Science and Technology Center, 320 Sparkman DR., 
SD50, Huntsville, AL 35805, USA}
}

\begin{abstract}
When thermal relativistic electrons with isotropic distribution of 
velocities move in a gas region, or impinge upon the surface of a 
cloud that consists of a dense gas or doped dusts, the Cerenkov effect 
produces peculiar atomic or ionic emission lines --- the Cerenkov 
line-like radiation. This newly recognized emission mechanism may find 
wide applications in high-energy astrophysics. 
In this paper, we tentatively adopt this new line emission
mechanism to discuss the origin of iron K$_{\alpha}$ feature
of AGNs. Motivation of this research is to attempt a solution to a 
problem encountered by the ``disk-fluorescence line'' 
model, i.e. the lack of temporal response of the observed 
iron K$_{\alpha}$ line flux to the changes of the X-ray 
continuum flux. If the Cerenkov line emission is indeed 
responsible significantly for the iron K$_{\alpha}$ feature,
the conventional scenario around the central supermassive 
black holes of AGNs would need to be modified to accommodate 
more energetic, more violent and much denser environments than 
previously thought.
\end{abstract}


%%%%%%%%%%%%%%%%%%%%%%%%%%
\newpage
\section{Introduction}

Observations in the last decade show that many 
active galactic nuclei(AGNs), e.g. the Seyfert 1 galaxies, 
display in their spectra an emission feature 
peaked around $\sim 6.4-6.5 {\mathrm KeV}$, commonly attributed to 
the K$_{\alpha}$ line emission of iron ions in low- or 
intermediate-ionization states. The observed K$_{\alpha}$ line is 
very broad, and the line profile is asymmetric, being steep
on the blue and flattening on the red wavelength wing, extending 
to $4-5 {\mathrm KeV}$, as shown in Fig.~1 (Tanaka et al. 1995; 
Nandra et al. 1997a, 1997b; Fabian et al. 2002; Wang et al. 2001).
The iron K$_{\alpha}$ is regarded as one of  the  best probes 
to explore the physical mystery in regions proximate to 
the central supermassive black holes of AGNs.  Its observation and 
interpretation thus have drawn great attention lately in black hole 
and AGN study.\\
% \put(100,20){\frame{Editor: please place Fig. 1 here}}

So far the prevailing model is based on the 
``photoelectric absorption-fluorescence line emission'' 
mechanism (e.g.,Guilbert et al. 1998; Lightman et al. 1998; 
Fabian et al. 1989; Reynolds 2001), which has gained wide popularity 
because it successfully produces a line-profile consistent 
with observations. Furthermore, the underlying emission 
mechanism, that is, photoelectric absorption followed by 
fluorescence line emission, has been so far taken for granted 
as the only way to produce the X-ray atomic or ionic emission 
line by heavy ions in low- or intermediate-ionization states for 
which the K-shell of ion is fully filled. Take an iron ion as 
an example, because the K-shell is fully closed, so the transition 
$n= 2 \rightarrow 1$ (K$_{\alpha}$) cannot occur unless certain 
external X-ray illumination causes photoelectric absorption to 
first make a  ``vacancy'' in the K-shell.

Despite the success of the ``photoelectric absorption-fluorescence 
line'' mechanism, we suggest an alternative mechanism---the 
Cerenkov line-like radiation---to explore the origin of iron
K$_{\alpha}$ in AGNs. The motivation of this research is to 
attempt a solution of a problem encountered by the ``disk-fluorescence line'' 
model, i.e. the temporal response of the iron K$_{\alpha}$ line
flux to the changes of the X-ray continuum flux, predicted by a simple
``photoelectric absorption-fluorescence line emission'' model. So far 
no clear response of line flux to the incident X rays has been 
observed (e. g. Lee et al. 1999, 2000; Chiang et al. 2000;
Wang et al. 1999, 2001; Weaver et al. 2001).  It has been suggested 
that a flux-correlated change of ionization states of 
the iron ions would be responsible for the lack of 
correlation between the fluxes of iron K$_{\alpha}$ line 
and the continuum (e.g. Reynolds  2001). This may well be true, 
and deserves to give a further quantitative analysis to confirm 
this viewpoint. There is another model that attempts 
to explain the lack of temporal response. Vaughan \& Fabian 
(2002) and Miniutti et al. (2003) suggest a model in which 
the X-ray source is close to the spin axis of the black hole,
 and the long-time scale changes ($>10$ ks) are due to 
changing height of this source. The light bending effect 
can then produce an almost constant line intensity together 
with a changing observed continuum flux.
In this paper, we try to give another
explanation for this problem by use of the newly recognized 
Cerenkov line emission mechanism.

In sec. 2 of this paper, we first outline the physics of the new 
line emission mechanism to help people who are unfamiliar 
with this mechanism. Relevant basic formulae are presented 
in Appendix. Besides, we discuss the conditions under which 
the new emission mechanism become predominant over the 
photoionization-fluorescence process, and should be taken into 
consideration to explore the origin of the iron K$_{\alpha}$ line 
in AGNs. In sec. 3 we give some model considerations and model 
calculations by use of the new mechanism to match the observed 
luminosities of the iron K$_{\alpha}$ line, and in turn to see whether 
the estimated environmental parameters are reasonable and 
acceptable in regions proximate to the central supermassive 
black hole of AGNs. Sec. 4 is conclusions and discussions.

\section{Cerenkov line-like radiation as the responsible 
mechanism---Outline of physics of the new mechanism}

In this paper we propose that the ``Cerenkov line-like radiation'' 
(You et al. 2000, 1980, 1986) could be responsible for the iron 
emission feature in AGNs. We shall show that this emission 
mechanism may become predominant over the fluorescence process 
under certain conditions around AGNs. Furthermore, the possible 
difficulty encountered by the fluorescence mechanism as mentioned 
above, namely the lack of correlation between the line and the 
continuum fluxes, can be alleviated because the radiation energy of 
Cerenkov line is provided by relativistic electrons rather than by 
the X-ray continuum as in fluorescence process.

The Cerenkov line-like mechanism has been confirmed by elegant 
laboratory experiments in O$_{2}$, Br$_{2}$ and Na vapor using 
a $^{90}$Sr $\beta$-ray source with the fast coincidence technique 
(Xu et al. 1981, 1988, 1989). Detailed discussions on the basic 
physics and improved formulae have been further presented recently 
(You et al. 2000, hereafter Y00). Here we outline the physics and 
essential results of the theory to help the people who are unfamiliar 
with this new mechanism. The relevant basic formulae are presented 
in Appendix of this paper for people who are interested in the theory 
of Cerenkov line-like radiation. 

When the thermal relativistic electrons with isotropic distribution of 
velocities move in a gas region, or impinge upon the surface of a dense 
cloud with arbitrary shape (e.g. with filamentary or sheet-like structure), 
Cerenkov radiation is produced within a narrow wavelength range 
$\Delta \lambda$, very close to the intrinsic atomic or molecular wavelength 
$\lambda_{lu}$ ($u$ and $l$ denoting respectively the corresponding upper 
and lower energy levels) because only in this narrow band the refractive 
index of gas is significantly larger than unit, $n>1$, which makes it 
possible to satisfy the Cerenkov radiation condition 
$n \geq \frac{c}{v}\equiv\frac{1}{\beta}$. The emission feature therefore 
appears more like an atomic or molecular line than a continuum, thus the name 
``Cerenkov line-like radiation'', or simply ``Cerenkov emission line''. 

For gaseous medium, the dispersion curve $n\sim \lambda$ and the resonant 
line-absorption curve $\kappa \sim \lambda$ can be calculated exactly 
by use of the formula of the refractive index for gas,  
$\frac{\tilde{n}^{2}-1}{\tilde{n}^{2}+2}=\frac{4\pi}{3}N\alpha$,
where $\tilde{n}=n-i\kappa$ is the complex refractive index, 
with the real part $n$ being the refractive index of gas, and the imaginary part 
$\kappa$ being the extinction coefficient which relates with the line-absorption 
coefficient by a simple formula 
$k_{lu}=\frac{4\pi \nu}{c}\kappa_{\nu}$ (Y00, or eq. (13) in Appendix);  
$N$ is the number density of the atomic/molecular species; $\alpha$ is the 
polarizability per atom or ion, given by quantum theory (Y00). 
For a very dense gas, $n$ is large. For 
$\lambda \approx \lambda_{lu}$, the value of $\alpha$, and hence the value 
of $n$, becomes very large 
(see the schematic dispersion curve $n^{2}\sim \lambda$ in FiG. 2). 
We emphasize that $n>1$ at $\lambda~^{>}_{\sim} \lambda_{lu}$ validates even when 
the lower energy level $l$ is fully closed, as long as the 
upper level $u$ is not completely filled (see Eq. (11) in Appendix). 
This is substantially different with 
the fluorescence emission, which always requires to preempt a vacancy in the 
lower level. It is this unique property that makes it easier to produce 
the K$_{\alpha}$ line of iron ions in intermediate-ionization states by the 
Cerenkov mechanism, under certain circumstances, than by the 
fluorescence process.\\
% \put(100,20){\frame{Editor: please place Fig. 2 here}}

In FiG.2 we see a strong resonant absorption occurs at $\lambda =\lambda_{lu}$ 
where the Cerenkov radiation vanishes. The Cerenkov mechanism only 
operates in the narrow shaded region $\lambda > \lambda_{lu}$, where the 
absorption approaches zero, $\kappa_{\lambda}\rightarrow 0$. The combination 
of absorption and emission causes the final emission feature slightly 
redshifted, which we call ``Cerenkov line redshift'' in order to distinguish 
it from other types of redshift mechanisms (Doppler, gravitational, 
Compton, etc.). As we shall 
show below, the Cerenkov line redshift is favorable to increase the 
emergent flux of Cerenkov emission line from the surface of a dense cloud.

In summary, the Cerenkov line-like emission has the following characteristics: 
(1) It is concentrated in a small wavelength range, so appears more like a line 
than continuum. The denser the gas, the broader the emission `line' feature. 
(2) If the dense gas is optically thick for the Cerenkov line emission, 
the emergent line profile becomes asymmetric, 
being steep on the high energy side and 
flattened on the low energy side. (3) The peak of the emission feature is not 
exactly at $\lambda =\lambda_{lu}$ but slightly redshifted due to the line 
absorption shown in FiG.2. In the optically thick case, 
the typical value of  `Cerenkov line redshift' would be so high as  
$z \sim 10^{-3}$ (Y00), which in terms of Doppler effect would correspond 
to an apparent velocity of several hundred kilometers per second. 
(4) The radiation would be polarized if the relativistic 
electrons have an anisotropic velocity distribution. 

FiG.3 shows the calculated 
profile of the Cerenkov K$_{\alpha}$ line of  Fe$^{+21}$ 
in optically thick case. For comparison a normal 
line by a spontaneous transition $n=2\rightarrow 1$ of Fe$^{+21}$ ion is also shown
in FiG.3 . The differences are obvious. \\
% \put(100,20){\frame{Editor: please place Fig. 3 here}}

The redshift effect(item 3 above) conveniently provides 
a mechanism in favor of the emergence of Cerenkov line 
emission, particularly from dense clouds. Obviously, 
for an opaque, optically thick dense gas, the emergent 
line flux from the surface of cloud is determined by the 
competition between emission and absorption. The 
absorption mechanism for a Cerenkov line is drastically 
different from that for a normal line. A normal spectral 
line, exactly located at $\lambda =\lambda_{lu}$, would be 
greatly weakened by a strong resonant line-absorption because 
the line absorption coefficient $k_{lu}(\lambda=\lambda_{lu})$ 
at $\lambda =\lambda_{lu}$, is very large(Fig. 2). In case of 
a very dense gas, the emergent radiation simply becomes a 
black-body continuum, and the normal line vanishes. In contrast, 
a Cerenkov line, occurs at $\lambda >\lambda_{lu}$ owing to 
the Cerenkov redshift, can avoid the strong line absorption because
$k_{lu}(\lambda>\lambda_{lu})\rightarrow 0$ (FiG. 2). Therefore 
a Cerenkov line suffers only very small amounts of photoelectric 
absorption $k_{\mathrm bf}$ (the extremely weak free-free absorption
$k_{\mathrm ff}$ in X-ray band can be neglected), much smaller 
than the regular line absorption $k_{lu}(\lambda_{lu})$, i. e.  
$k_{\mathrm bf}\ll k_{lu}(\lambda=\lambda_{lu})$. For example, 
for the iron K$_{\alpha}$ line, the dominant photoelectric absorption
comes from  the L-shell electrons of iron ions, for which the 
photoelectric absorption coefficient 
$k_{\mathrm bf}\approx k_{\mathrm bf}({\mathrm {Fe, L}})$ is much 
smaller than the regular line absorption. That is,
$k_{\mathrm bf}\approx  k_{\mathrm bf}({\mathrm {Fe, L}})\ll 
k_{lu}(\lambda =\lambda_{lu})$. 
This means that the photons of Cerenkov line can escape readily from 
deep inside a dense gas cloud; in other words,
the dense gas would appear more `transparent' for 
the Cerenkov line emission than for a normal line 
produced by the spontaneous transition. It is probable 
that the optical depth of a dense cloudlet with size $r$ 
can be less than unit, $\tau=k_{\mathrm bf}({\mathrm {Fe, L}})r<1$,
despite of the high density of iron ions $N_{\mathrm Fe}$, i. e. 
the dense cloudlet possibly becomes optically thin for the peculiar 
Cerenkov line emission. Even if in the optically thick case, 
$\tau=k_{\mathrm bf}({\mathrm {Fe, L}})r>1$, 
the Cerenkov emission layer at the surface of dense cloud 
with thickness $l\sim 1/k_{\mathrm bf}({\mathrm {Fe, L}})$
would still be surprisingly thicker than that for normal line.
An optically thin case or a thick Cerenkov emission layer 
at the surface of an opaque dense gas region means 
a possibility of very strong emergent Cerenkov line emission, 
as long as there are sufficient number of relativistic electrons 
near the surface. It is possible that the Cerenkov line emission 
is even predominant over the markedly suppressed normal fluorescence 
line in such special cases.

\section{Model considerations and calculations}
{\bf 3.1. Model considerations --- new scenario}~~~As mentioned above, The Cerenkov line-like radiation may be 
particularly important in astrophysical environments with 
a very high gas concentration and with abundant relativistic 
electrons. One such example would be AGNs, particularly the 
Seyfert 1 galaxies, for which the existence of dense gas 
region seems plausible. The gas at the surface of an AGN disk
is thought to be compressed to very high density 
by the high radiation pressure of the coronal X rays.
However, although the disk-type geometry is 
compatible with the Cerenkov mechanism, in the 
following model consideration, we prefer to adopt 
the quasi-spherical distribution of dense cloudlets 
with spherical, filamentary or sheet-like shapes 
around the central black hole, in order to avoid 
some defiances on the validity of disk models 
(Sulentic et al. 1998a;1998b). The possible 
presence of such dense clouds, filaments and 
sheets in AGN environments has been discussed 
by Rees(1987);Celotti, Fabian \& Rees(1992); 
Kuncic, Blackman \& Rees (1996); Kuncic, 
Celotti \& Rees (1997) and Malzac (2001). 
The clouds must be very dense to remain cool and 
therefore held by magnetic fields. Cool gas trapped 
by the magnetic field is compressed to extreme densities 
by the high radiation pressure, as what happens at the 
surface of disk surrounding the central supermassive 
black hole. Recently some authors adopt the quasi-spherical 
distribution of dense cloudlets to explain the origin and 
profile of the iron K$_{\alpha}$ line ( Karas et al. 2001; 
Collin-Souffrin et al. 1996; Brandt et al. 2000). In their scenario, 
the innermost part of the disk is disrupted due to disk instabilities. 
Part of the disrupted material forms the optically thick, cold cloudlets 
of dense gas that cover a significant portion of the sky from the point 
of view of the central X-ray source, while the rest gets heated 
up to high temperatures, forming a corona (see FiG.4). Recent 
observations support the existence of dense clouds or filaments 
in AGNs (Boller et al. 2002) which strongly favors the operation 
of Cerenkov line-like radiation mechanism.

FiG.4 sketches the schematic of our working model 
of a quasi-spherical emission region around the central 
supermassive black hole of a AGN. In FiG.4 the shaded 
spots or stripes represent cool cloudlets or filaments 
of dense gas. The dotted region stands for the hot, 
rarefied corona. The tiny dots, uniformly distributed 
in corona, represent thermal electrons and the black 
dots represent the relativistic electrons, which are  
highly concentrated around the cloudlets in corona 
(reasons see below).\\
% \put(100,20){\frame{Editor: please place Fig. 4 here}}

Evidence also seems to be mounting on the existence of 
abundant relativistic electrons. It is likely that the observed 
power-law continuum over a very wide frequency range, 
from radio to UV is largely attributed to non-thermal 
radiation of relativistic electrons.\footnote{However, 
there is no clear evidence for X-ray emission 
from highly relativistic electrons in Seyfert galaxies.}
Although the detail mechanism 
to produce an excessive amount of high energy electrons 
remains unclear, flare events or some shock processes 
in corona may be responsible. Such shock processes also 
take place in the gamma-ray burst events, in which ultra-fast 
electrons are produced by the internal and external shock waves, 
thus producing the nonthermal radiation.
The strong shock waves originate from drastic 
release of gravitational energy during the mergers processes, 
e. g. the neutron star-neutron star or the neutron star-black 
hole mergers (Piran 1999, 2000; Meszaros 2002). 
It is probable that similar processes also occur 
in AGNs environments. The biggest difference 
between the accretions of the AGNs supermassive 
black hole and the small black hole with solar mass 
could be that the accreted matter moving around the 
AGNs black hole is not in a pure gas state. Many 
components could be coexistent and mixed in regions 
proximate to the central black hole. Except of the 
dense cloudlets, there exist the solid debris and fragments, 
the meteorites and planets, even the stars, e. g. the neutron 
stars and the black holes with star mass. The frequent 
collisions and mergers between these objects should be 
expected, e. g. the mergers of meteorite-neutron star and 
mergers of planet-black hole, etc., which also produce a 
chain of `merger-drastic release of gravitational energy-strong 
shock-plenty of fast electrons', though the scale of energy 
release in each merger could be much smaller than that in GRBs.
Given the ubiquity of shock events in corona region around the 
central supermassive black hole, and their frequent collisions 
with dense clouds, the whole region of iron line emission is 
regarded as a shock-filled X-ray source(FiG.4). The collisions 
convert part of the kinetic energy of shock wave to thermal 
energy of relativistic electrons (see, e. g. Piran 1999). Obviously most fast 
electrons thus produced are concentrated in a narrow zone 
which closes to the collision-front at the surface of the dense 
cloudlet (or on the flare spots if they are produced in flare events), 
as shown in FiG.4. Diffusion of fast electrons outward to ambient 
space is expected to be slow owing to the trapping effect of 
magnetic fields near the clouds or flares. The ability to retain 
extremely high concentration of relativistic electrons near 
the dense clouds undoubtedly provides a very favorable 
environment for the operation of the Cerenkov line emission.

If the conditions, namely the existence of dense gas and 
relativistic electrons, are met, the Cerenkov line like 
emission becomes inevitable. 

{\bf 3.2. Model calculation}~~~Now we derive the intensity or luminosity of the Cerenkov line, 
in particular the iron K$_{\alpha}$ line, under various environmental 
parameters, and compare it with observations. Except for the 
geometric size of emission region, the main factors that 
determine the luminosity of the Cerenkov line include the 
density of the iron ions $N_{\mathrm {Fe}}$, the density of 
the relativistic electrons $N_{\mathrm e}$, and the average or 
typical energy $\gamma_{\mathrm c}$ of the relativistic electrons, 
where the Lorentz factor 
$\gamma \equiv \frac{1}{\sqrt{1-\beta^{2}}}=\frac{mc^{2}}{m_0c^2}$ 
represents the dimensionless energy of an 
electron in unit of  $m_{0}c^{2}$. Our goal is to estimate the 
(range of ) values of $N_{\mathrm e}$, $N_{\mathrm Fe}$
and $\gamma_{\mathrm c}$ under a fixed size $D$ of emission region
from the theory of Cerenkov line radiation(Y00), to match the 
observed luminosities of the iron K$_{\alpha}$ line, and in turn to 
see whether these values are acceptable in the environment of AGNs.

In this paper, our main interest is in the energetics of the iron 
K$_{\alpha}$ emission process, therefore we leave out lengthy 
discussion on the line profile, except to note that even though 
a Cerenkov line is intrinsically broad, asymmetric and redshifted, 
it is still insufficient to produce the highly skew emission features 
observed in AGNs. A supermassive black hole is still needed to 
provide the necessary Doppler broadening and gravitational redshift.

We assume a typical mass $M\sim 10^{7-8} M_{\odot}$ for the central 
black hole. From the observed variation time scales 
of the iron K$_{\alpha}$ line  ($\sim$ 1 lt day, see  Nandra, George, Mushotzky, 
et al. 1997; Done et al. 2000), we infer the size of 
the emission region to be $D \sim 10^{15-16} {\mathrm cm}$. 
In the following model calculations, we adopt a typical value 
$D\approx 3\times 10^{15} {\mathrm cm}$, 
or about $\sim 10^{2-3} R_{\mathrm Sch.}$, 
where $R_{\mathrm Sch.}$ is the Schwarzschild radius. To simplify 
the calculation, we assume all of the dense gas regions in the 
form of spherical cloudlets with the same radius, except to note that 
in reality they will naturally be in various cloud-like, filament-like 
or sheet-like shapes with various sizes. As noted before, because of 
the strong irradiation from the central X-ray source, there should exist 
a photo-ionized layer at the surface of a cloudlet. The main species 
of the iron ions in this layer should be in the 
intermediate-ionization states, e.g. those from Fe$^{+18}$ to 
$\sim$ Fe$^{+21}$, because in many cases the observed line-centers 
are around $\sim 6.47-6.5\mathrm KeV$ (e.g. Wang et al. 1998; 
Weaver et al. 2001). Denoting the total number of cloudlets in the 
whole emission region as $\widetilde{N}$, and the radius of each 
cloudlet as $r$, therefore the covering factor of 
cloudlets to the central X-ray source is 
\begin{eqnarray}
f_{\mathrm c}=\frac{\widetilde{N}
\pi r^2}{4\pi D^2}=\frac{\widetilde{N}r^2}{4D^2},
\end{eqnarray}
which must be less than unity to ensure the central X-ray source 
to be only partly covered. 

From Eq.(1) we see that the unknown 
quantity $\widetilde{N}r^{2}$ can be removed by $f_{\mathrm c}$ 
and $D$. The volume filling factor is 
\begin{eqnarray}
f_{\mathrm v}=\frac{\widetilde{N}4\pi r^3/3}{4\pi D^3/3}=
\widetilde{N}\frac{r^3}{D^3}.
\end{eqnarray}
Therefore the ratio is $f_{\mathrm v}/f_{\mathrm c} \approx r/D$, 
which means that the fractional volume occupied by the 
cloudlets would be very small if $r \ll D$. At the same time,  
$\widetilde{N}$  can be still very large to maintain a 
necessarily large covering factor $f_{\mathrm c}$.

In order to avoid the ambiguity of our quantitative analysis, 
the following is restricted to calculate the Cerenkov line 
emission in the optically thick case, though the optically 
thin case is also  possible in AGNs.

The emergent intensity of Cerenkov iron K$_{\alpha}$ line from 
the surface of the optically thick dense gas
is (Eq. (22) in Appendix, or Eq.(42) in Y00)
\begin{eqnarray}
I_{K_{\alpha}}^{\mathrm c}=Y\left [ \ln (1+X^{2})-
2\left (1-\frac{\arctan X}{X}\right )\right ]
~~{\mathrm (ergs/Sec. \cdot cm^{2}\cdot Str.)},
\end{eqnarray}
where the parameter $Y \equiv \frac{N_{\mathrm e}C_{1}}{2 k_{\mathrm bf}}
\propto N_{\mathrm e}$, the density of relativistic 
electrons;  and $X \equiv \sqrt{\frac{k_{\mathrm bf}}{C_{2}}} 
C_{0} \gamma_{\mathrm c}^{2}\propto  \gamma_{\mathrm c}^{2} N_{\mathrm Fe}$,  
where $N_{\mathrm Fe}$  and $\gamma_{\mathrm c}$ represent the density of 
iron ions in gas cloudlets and the typical (or average) energy of the 
relativistic electrons, respectively. $C_{0}$,  $C_{1}$, $C_{2}$ and 
$k_{\mathrm {bf}}\approx k_{\mathrm {bf}}(\mathrm {Fe, L})$ 
included in $X$ and $Y$, are the parameters which are dependent on 
the density $N_{\mathrm Fe}$ as well as the atomic parameters of  concerned ions, 
e.g. the frequency $\nu_{lu}$ (or $h\nu_{lu}\equiv \varepsilon_{lu}$), 
the transition probability $A_{ul}$, etc. 
(Eq. (21) in Appendix or Y00). Inserting the concerned 
atomic parameters of the iron ions, for iron K$_{\alpha}$ line, 
$u=2$, $l=1$, we obtain 
\begin{eqnarray}
&& X \equiv \sqrt{\frac{k_{\mathrm bf}}{C_{2}}} C_{0} \gamma_{\mathrm c}^{2}
=6.49\times 10^{-28}g_{2}\sqrt{\frac{S_{2}}{g_{2}}\left (\frac{S_{1}}{g_{1}}
-\frac{S_2}{g_2}\right )}N_{\mathrm Fe} \gamma_{\mathrm c}^{2},\nonumber \\
&& Y\equiv \frac{N_{\mathrm e}C_{1}}{2 k_{\mathrm bf}}=0.16\times 
\frac{g_{2}}{S_{2}}\left (\frac{S_{1}}{g_{1}}-\frac{S_{2}}{g_{2}}\right )N_{\mathrm e} ,
\end{eqnarray}
where $g_{2}$ and $S_{2}$ respectively represent the degeneracy and 
the real occupation number of electrons at the second level of 
the iron ion, so $S_{2}\leq g_{2}$. $g_{1}$ and $S_{1}$ are 
the corresponding quantities of the first level.

From Eq.(4) we see that in a physically reasonable environment in AGNs, 
it usually holds that $X< 1$. Therefore Eq.(3) is simplified as  
\begin{eqnarray}
I_{K_{\alpha}}^{\mathrm c}\approx Y X^{2}/3,
\end{eqnarray}

The  Cerenkov line emission produced by the thermal relativistic electrons 
with random direction-distribution of velocities is isotropic, 
thus the Cerenkov intensity $I_{K_{\alpha}}^{\mathrm c}$ is 
$\theta$-independent. Therefore the emergent flux $F_{K_{\alpha}}^{\mathrm c}$ 
from the surface of the dense cloud is simply obtained
\begin{eqnarray}
F_{K_{\alpha}}^{\mathrm c}=2\pi \int_{0}^{\pi/2} I_{K_{\alpha}}^{\mathrm c} 
\cos \theta \sin \theta {\mathrm d} \theta
=\pi I_{K_{\alpha}}^{\mathrm c} ~~{\mathrm  (ergs/Sec.\cdot cm^{2})}.
\end{eqnarray}

So the elementary luminosity of the Cerenkov iron K$_{\alpha}$ 
line of each cloudlet is
\begin{eqnarray}
l_{K_{\alpha}}^{\mathrm c}=4\pi r^{2} F_{K_{\alpha}}^{\mathrm c}
=4\pi^{2} r^{2} I_{K_{\alpha}}^{\mathrm c}
~~{\mathrm (ergs/Sec.)}.
\end{eqnarray}

Combining Eq.(1) and (7), the total luminosity of Cerenkov iron 
K$_{\alpha}$ line from the whole emission region becomes 
\begin{eqnarray}
L_{K_{\alpha}}^{\mathrm c}= \widetilde{N}l_{K_{\alpha}}^{\mathrm c}
=4\pi^{2} r^{2}\widetilde{N}I_{K_{\alpha}}^{\mathrm c}\approx 
16 \pi^{2} D^{2} f_{\mathrm c} I_{K_{\alpha}}^{\mathrm c}
~~{\mathrm (ergs/Sec.)}.
\end{eqnarray}

Inserting Eq.(4),  (5) into Eq.(8), taking $D\approx 3\times 10^{15} {\mathrm cm}$ 
and $f_{\mathrm c}\approx 0.1$, and $S_{1}=g_{1}=2$, we obtain
\begin{eqnarray}
L_{K_{\alpha}}^{c}=2.08\times 10^{-22}\left ( 1-\frac{S_2}{g_2}\right )^{2} 
N_{e}N_{\mathrm {Fe}}^{2}\gamma_{c}^{4} ~~\mathrm (ergs/Sec.),
\end{eqnarray}

Comparing the Cerenkov luminosity of iron K$_{\alpha}$ line Eq.(9) with 
the typical observed value for Seyfert 1s, i. e.
$L_{K_{\alpha}}^{\mathrm obs.}\approx 10^{40-41} ~{\mathrm ergs/sec.}$, letting 
$L_{K_{\alpha}}^{c}\approx L_{K_{\alpha}}^{\mathrm obs.}$,  we obtain
\begin{eqnarray}
N_{\mathrm {Fe}}^{2}\gamma_{c}^{4}N_e \approx 4.8\times 10^{61} 
\left (1-\frac{S_2}{g_2} \right )^{-2}\approx 10^{62},  
\end{eqnarray}
where for iron ions in intermediate-ionization states, we take 
$S_{2}\approx 1-5$, which corresponds to iron ions Fe$^{+19}$---Fe$^{+23}$.  
Eq.(10) gives the combination condition for the iron density 
$N_{\mathrm {Fe}}$,  the density of fast electrons $N_{\mathrm e}$, 
and the average or typical energy $\gamma_{\mathrm c}$ of 
the fast electrons to produce the Cerenkov luminosity of 
iron K$_{\alpha}$ line which can be compared with the observed value.

Table 1 lists several tentative sets of parameters under the condition, 
where we arbitrarily fix $N_{e}=10^{10}~{\mathrm cm^{-3}}$. The choice 
of combinations is somewhat arbitrary, because so far the environments 
in AGNs are still not well understood. Some of these parameters may at first 
appear defiant to the current paradigm around the AGN black holes. 
In the following section we give discussions on the reasonableness 
of these parameters and acceptability of the related new scenario around 
the supermassive black hole in AGNs. \\
% \put(100,20){\frame{Editor: please place Table 1 here}}

\section{Conclusions and Discussions}

4.1.In this paper, we tentatively propose another 
mechanism---the ``Cerenkov line-like radiation''---to study 
the origin of the elusive iron K$_{\alpha}$ feature in AGNs. 
The charming advantage of this new emission mechanism is 
that the radiation energy of the Cerenkov line is provided 
by the relativistic electrons rather than by the X-ray continuum. 
Therefore the continuum and the iron K$_{\alpha}$ line emission 
are two components independent of each other. So the lack 
of correlation between the line and the continuum fluxes can be 
understood by this way. We further give some model calculations
to show the effectiveness of the Cerenkov mechanism to explain 
the AGNs observations. We show that the calculated Cerenkov
iron K$_{\alpha}$ line is strong enough to compare with 
observations of AGNs, only if the iron density $N_{\mathrm Fe}$  
in the dense gas, the density of fast electrons $N_{\mathrm e}$
and the characteristic energy of fast electrons $\gamma_{\mathrm c}$ 
are high enough, as shown in Table 1.

If the iron K$_{\alpha}$ feature indeed arises from the Cerenkov 
line mechanism, Table.1 signifies the possibility of a strikingly different 
scenario around the supermassive black hole in AGN---with much denser, 
more violent and more energetic environs than conventionally believed. 
In the following 4.2-4.4 sections, we discuss the reasonableness and 
acceptability of the new scenario.

4.2. Firstly, what are the consequences if there exist abundant 
relativistic electrons with exceedingly high energies?  The problem 
is in that, except for the Cerenkov line-like radiation, the fast electrons 
also contribute substantially to the continuum radiation in high energy 
band through the inverse Compton scattering process (the synchrotron 
mechanism becomes unimportant in a very dense gas). The Compton 
power of a fast electron with energy $\gamma_{\mathrm c}$ 
passing through an X-rays field with energy density $U_{\mathrm ph}$
is as high as
$p^{\mathrm Comp.}\approx 2.6\times 10^{-14}U_{\mathrm ph}
\gamma_{\mathrm c}^{2}\approx 10^{-15}\gamma_{\mathrm c}^{2}$ 
($U_{ph}\approx 0.1~{\mathrm ergs/cm^{3}}$ for typical Seyfert 1 galaxies).
If the density and the energy of the fast electrons are so high as 
$N_{\mathrm e}\approx 10^{9-10}{\mathrm cm^{-3}}$, 
and $\gamma_{\mathrm c}\approx 10^{4-5}$, as shown in Table 1, the Compton 
luminosity of the continuum in the whole emission region with size 
$D\sim 10^{15}{\mathrm cm}$ would be unacceptably higher than the typical 
observed value $L \approx 10^{44} {\mathrm ergs/Sec.}$. 
Another related problem is that, if both the iron K-line emission 
and a significant portion of the continuum radiation owing to 
the inverse Compton scattering process originated from the same 
group of fast electrons, again the correlation between the fluxes 
of iron K-line and the X-ray continuum would be expected, as in the 
case of fluorescence model. 

We envisage a solution to overcome the difficulties mentioned above. 
As pointed out in sec. 3, the fractional volume occupied by the cloudlets, 
where most of fast electrons reside, can be very small compared with the 
overall volume of X-ray emission region $V=\frac{4\pi}{3}D^{3}$,  
i.e. $f_{\mathrm v}\ll 1$. In this case, the total number of fast 
electrons may not be large, thus the corresponding Compton 
luminosity of the continuum $L^{\mathrm Comp.}\approx f_{\mathrm v}
\left (\frac{4\pi}{3}D^{3}\right )N_{e} p^{\mathrm Comp.}$ 
would be several orders lower 
than the typical value of the observed continuum luminosity 
$L \approx 10^{43-44} {\mathrm ergs/Sec.}$. A small $f_{\mathrm v}$
with a large $f_{\mathrm c}$ is readily realized as long as
$r\ll D$, as mentioned in sec. 3. For example, taking $D\sim 10^{15}~{\mathrm cm}$,
$r\sim 10^{6-7}~{\mathrm cm}$, and $f_{\mathrm c}\sim  0.1$, then the filling 
factor is small as $f_{\mathrm v}\sim 10^{-10}-10^{-9}$, hence we get 
$L^{\mathrm Comp.}\ll L^{\mathrm obs.}\sim 10^{44}~{\mathrm ergs/Sec.}$.

4.3.  Another potential problem concerns the very high density of gas.  
If $N_{\mathrm Fe}$ is so high as those shown in Table 1,
even up to $\sim 10^{17-18}~{\mathrm cm^{-3}}$, and if a cosmological 
abundance in AGNs is assumed, then the inferred gas density 
would be inconceivably high. Therefore an abnormally 
high iron abundance is necessary to restrict the total gas density at an 
acceptable level. We envisage two possibilities as follows: either 
frequent nuclear reactions in the vicinity of the central black hole 
or a phase transition to form dusty clouds in dense gas regions, could be 
responsible for the marked increase of abundance of iron and 
other heavy elements without enhancement of gas density. 
In the later case, the iron ions may be locked up in tiny grains 
as embedded impurity. A high density of impurity iron is achievable 
in a heavily doped solid, even as high as 
$N_{\mathrm Fe}\sim 10^{17-18}~{\mathrm cm^{-3}}$. 
It is conceivable, in principle, that the Cerenkov line-like radiation 
of iron ions may also occur in the impurity-doped dust, as in gas medium. 
Undoubtedly, the Cerenkov line-like radiation from the impurity-doped 
solid would be a challenging problem in experimental physics in future. 
Existence of dusty clouds with iron-rich grains in the environments of 
AGNs would be equally mind-boggling in black hole physics.

4.4. Finally, we concern the reasonableness of the energetics of our model. 
As we suggested in sec.3, the relativistic electrons, necessary for the Cerenkov 
line-like radiation mechanism, are produced, for example, by the strong shocks,
which originate from the mergers processes.  
However, the energy that goes into the relativistic electron population in this 
way could only be a small fraction of the total. In this case, it would be 
important to give a more comprehensive physical consideration on the 
`efficiency of energy transformation from the kinetic energy of 
shock to the thermal energy of relativistic electrons'. 
We hope to give a moderate solution in future.

\acknowledgements {We are  sincerely grateful to Dr. Christopher S. Reynolds 
in University of Maryland for his helpful discussions and suggestions, 
which help us to greatly improve this paper. 
The work of JHY is supported by the Natural Science Foundation of China, 
grant No.~19773005. WPC acknowledges the grant NSC91-2112-M-008-043 
from the National Science Council.}

%%%%%%%%%%%%%%%%%%%%%%%%%%%%%%%%%%%%%%%%%%%%%%%%%%%%%%%%%%%%%%%%%%%%%%%%%%%

%%%%%%%%%%%%%%%%%%%%%%%%%%%%%%%%%%%%%%%%%%%%%%%%%%%%%%%%%%%%%
\newpage
\section*{Appendix: Basic Formulae for Cerenkov line-like
radiation}

We first emphasize that the CGSE system of units was used 
in the Appendix. This means, in particular, that all energies 
of X-ray photons in the following formulae will be in ergs 
rather than KeV (1~KeV=1.602$\times 10^{-9}$~erg). Besides, one can find the detailed derivation 
of the following basic formulae in our published paper (You et al. 2000). 

\subsection*{1. \, The refractive index $n$ and the
 extinction coefficient $\kappa$}

The essential point of the calculation of the spectrum of Cerenkov
radiation is the evaluation of the refractive index of the gaseous
medium. This is easy to understand qualitatively from the
necessary condition for producing Cerenkov radiation, $v>c/n_\nu$.
At a given frequency $\nu$, the larger the index $n_{\nu}$, the
easier the condition $v>c/n_{\nu}$ to be satisfied, and the stronger
the Cerenkov radiation at $\nu$ will be. Therefore, in order to
get the theoretical spectrum of Cerenkov radiation, it is
necessary to calculate the refractive index $n_\nu$ and its
dependence on $\nu$ (the dispersion curve $n_\nu\sim\nu$). For
a gaseous medium, the calculation is easy to do. Omitting the
detailed derivation, here we only give the expressions of the index $n_\nu$ and the
extinction coefficient $\kappa_\nu$ of gas as follows
\begin{eqnarray} 
&&n_\nu^2-1=\frac{C^3h^4}{16\pi^3}\varepsilon_{lu}^{-4}
    A_{ul}g_uN_{\mathrm {Fe}}\left (\frac{S_l}{g_l}-\frac{S_u}{g_u}\right )y^{-1}
    \nonumber
\\ && \kappa_{\nu}=\frac{C^3h^4}{128\pi^4}\varepsilon_{lu}^{-5}
    \Gamma_{lu}A_{ul}g_uN_{\mathrm {Fe}}\left (\frac{S_l}{g_l}-\frac{S_u}{g_u}\right )y^{-2}
\qquad  ({\mathrm when } ~~y\geq 10^{-5})
\end{eqnarray}
where $\varepsilon_{ul}\equiv h\nu_{ul}= \varepsilon_u-\varepsilon_l$
represents the energy of line photon, $\varepsilon_u$ and $\varepsilon_l$
are the energy 
of the upper and lower levels of the iron ion, respectively 
(for definiteness, in the following, we only concern with 
the Cerenkov iron line, particularly the iron K$_{\alpha}$). 
$A_{ul}$ is the
Einstein's spontaneous emission coefficient for $u\rightarrow l$.
$\Gamma_{ul}=\Gamma_{u}+\Gamma_{l}=\sum_{i<u}A_{ui}+\sum_{j<l}A_{lj}$
is the quantum damping constant for the atomic(ionic) line with
energy $\varepsilon_{ul}$, which is related with the Einstein's
spontaneous emission probabilities $A_{ui}$ and $A_{lj}$.
$N_{\mathrm {Fe}}$ is the number density of iron ions in gas. $g_u$ (or
$g_l$), and $S_u$ (or $S_l$) are the degeneracy and the actual
occupation number of electrons of the upper level u (or lower
level $l$), respectively. $y\equiv\frac{\Delta\lambda}{\lambda_{ul}}
=-\frac{\Delta\nu}{\nu_{ul}}=-\frac{\Delta\varepsilon}{\varepsilon_{ul}}
=\frac{\varepsilon_{lu}-\varepsilon}{\varepsilon_{lu}}$ represents the
fractional displacement of the frequency or photon-energy. $y\ll 1$
owing to the fact that Cerenkov line-like radiation concentrates
in a narrow band $\varepsilon\approx \varepsilon_{ul}$.

\subsection*{2.\, The Cerenkov spectral emissivity $J_\nu^c$ (or
$J_\varepsilon^c$)}

The Cerenkov spectral emissivity can be derived from the
dispersion curve $n_\nu\sim\nu$ given above. It is known
from the basic theory of Cerenkov radiation that the power emitted in a
frequency interval $(\nu, \nu+{\mathrm d}\nu)$ or $(\varepsilon,
\varepsilon+d\varepsilon)$ by an electron moving with velocity
$\beta=\frac{v}{c}$ is $P_\nu d\nu=(4\pi^2e^2\beta
\nu/c)(1-\frac{1}{n_\nu^2\beta^2})d\nu$. Let $N(\gamma)d\gamma$ to be
the number density of fast electrons in the energy interval
$(\gamma, \gamma+d\gamma)$
($\gamma\equiv\frac{1}{\sqrt{1-\beta^2}}=\frac{mc^2}{m_0c^2}$ is the
Lorentz factor, representing the dimensionless energy of the
electron). For an isotropic velocity distribution of the
relativistic electrons as in normal astrophysical conditions, the
Cerenkov radiation will also be isotropic. Then the Cerenkov
spectral emissivity $J_\nu^cd\nu$, or equivalently,
$J_y^cdy=J_\nu^cd\nu$, can simply be obtained by the integral
$J_\nu^cd\nu=\frac{1}{4\pi}\int_{\gamma_1}^{\gamma_2}N(\gamma)d\gamma
P_\nu d\nu$, thus we get
\begin{equation}
J_y^c{\mathrm d}y=C_1N_e(y^{-1}-y_{\mathrm lim}^{-1}){\mathrm d}y
\end{equation}
where $N_{\mathrm e}$ is the density of fast electrons.
$y\equiv-\frac{\Delta\varepsilon}{\varepsilon_{ul}}$ is the fractional
displacement of energy $\varepsilon$ relative to intrinsic line-energy $\varepsilon_{ul}\equiv
h\nu_{ul}$. $y_{lim}=C_0\gamma_{\mathrm c}^2$ is the fractional
Cerenkov line-width.
$C_0$, $C_1$ are the coefficients which are dependent on the density
of iron ions $N_{\mathrm {Fe}}$ and the atomic parameters of concerned
species of iron ions, as shown below.

\subsection*{3.\, The absorption coefficient}
For an optically thick dense gas for which the Cerenkov line
mechanism is more efficient, the final emergent intensity $I_\nu^c$ is
dertermined by the competition between the emission $J_\nu^c$ (or
$J_\varepsilon^c$) and the absorption $k_\nu$. Therefore it is
necessary to consider the absorption of the gas at
$\nu\approx\nu_{lu}$. For the optical and X-ray bands, only two
absorption mechanisms are important for the Cerenkov line. One is
the line absorption $k_{lu}$ in the vicinities of atomic lines,
which directly related to the extinction coefficient $\kappa_\nu$
given in Eq. (11) by a simple 
relation, i. e. $k_{lu}=4\pi\nu\kappa_\nu/c$. Another is the
photoelectric absorption $k_{\mathrm {bf}}$. The free-free absorption 
in X-ray band is negligibly small. Thus the total absorption is
\begin{equation} 
k_{\nu}=k_{lu}+k_{\mathrm {bf}}=C_2y^{-2}+k_{\mathrm {bf}}
\end{equation}
where the coefficient $C_2$ depends on $N_{\mathrm {Fe}}$ and other atomic
parameters, as $C_0$, $C_1$ does. Owing
to the fact that the line absorption decreases with
$\Delta\nu=\nu-\nu_{lu}$ rapidly as $k_{lu}\propto y^{-2}$,
$k_{lu}(\nu<\nu_{lu})\rightarrow 0$. Therefore in the whole
actually effective frequency band of Cerenkov line emission, the dominant
absorption is $k_{\mathrm {bf}}$. Particularly, for iron $K\alpha$ line
which we concern, the dominant photoelectric absorbers are the
L-shell electrons of iron ions. In this
case,Eq.(13) becomes
\begin{equation}
k_{\nu}=k_{lu}+k_{\mathrm {bf}}\approx k_{\mathrm {bf}}\approx k_{\mathrm {bf}}(Fe, L)
\end{equation}
and
\begin{equation}
k_{\mathrm {bf}}(Fe, L)=N_{\mathrm {Fe}}S_2\sigma_{\mathrm {bf}}(2)
\end{equation}
where $S_2$ is the occupation number of electrons in L-shell, thus
$S_2\leq g_2=8$. $\sigma_{\mathrm {bf}}(2)$ is the cross section of
photoelectric absorption of a L-shell electron. For an iron atom
or ion, the hydrogen-like formula for the cross section is a good
approximation, particularly for the low-lying levels n=2, 3, i.e.
\begin{equation}
\sigma_{\mathrm {bf}}(\nu,n)=
    \frac{32\pi^2e^6R_{\infty}Z^4}{3\sqrt{3}h^3\nu^3n^5}g_{\mathrm {bf}}(\nu,T)
\end{equation}
Inserting to Eq.(15),
taking the effective charge number $Z^{\mathrm {eff.}}=24$ for the L-shell 
electrons at level $n=2$,
and let Gaunt factor $g_{\mathrm {bf}}\approx1$, we get
\begin{equation}
k_{\mathrm {bf}}({\mathrm Fe, L})\approx8.4\times10^{-46}N_{\mathrm {Fe}}S_2\varepsilon_{lu}^{-3}
\end{equation}

\subsection*{4.\, The emergent Cerenkov spectral intensity 
and Cerenkov total line intensity}

For an uniform plan-parallel slab, the emergent Cerenkov spectral
intensity $I_\nu^c$ from the surface of the slab can be derived by
use of the equation of radiative transfer
\begin{eqnarray}
I_\nu^c=\frac{J_\nu^c}{k_\nu}(1-e^{-k_\nu L})
%~~~~ {\mathrm or} ~~~~ I_y^c=\frac{J_y^c}{k_y}(1-e^{-k_y L})
\end{eqnarray}
where $J_\nu^c$ and $k_\nu$ ( or $J_y^c$ and $k_y$) are given by 
Eq. (12) and (13), respectively.

For optically thin gas, $\tau_{\nu}=k_{\nu}L\ll 1$, therefore the Cerenkov line intensity is 
\begin{eqnarray}
I_{\nu}^{c}\approx J_{\nu}^{c}L 
%~~~~~~{\mathrm or}~~~~~~  I_{y}^{c}\approx J_{y}^{c}L 
\end{eqnarray}

However, for a very dense gas, the Cerenkov emission 
slab becomes optically thick, thus Eq.(18) can
be simplified as $I_\nu^c\approx\frac{J_\nu^c}{k_\nu}$, or
equivalently,
\begin{equation}
I_y^c=\frac{J_y^c}{k_{lu}+k_{\mathrm {bf}}}
    =\frac{N_eC_1(y^{-1}-y_{lim}^{-1})}{C_2y^{-2}+k_{\mathrm {bf}}}
\end{equation}
where $y_{lim}=C_0\gamma_c^2$ is the fractional Cerenkov line width. All of
the atomic parameter-dependent coefficients $C_0$, $C_1$, $C_2$,
$k_{\mathrm {bf}}$ for iron ions are given as follows.
\begin{eqnarray}
&&C_0=1.05\times10^{-76}\varepsilon_{lu}^{-4}A_{ul}g_uN_{\mathrm {Fe}}
    \left (\frac{S_l}{g_l}-\frac{S_u}{g_u}\right ) \nonumber\\ 
&&C_1=5.77\times10^{-53}\varepsilon_{lu}^{-2}A_{ul}g_uN_{\mathrm {Fe}}
    \left (\frac{S_l}{g_l}-\frac{S_u}{g_u}\right ) \nonumber\\ 
&&C_2=1.75\times10^{-87}\varepsilon_{lu}^{-4}A_{ul}\Gamma_{lu}g_uN_{\mathrm {Fe}}
    \left (\frac{S_l}{g_l}-\frac{S_u}{g_u}\right ) \nonumber \\
&&k_{\mathrm {bf}}({\mathrm Fe, L})=8.4\times10^{-46}\varepsilon_{lu}^{-3}S_2N_{\mathrm {Fe}}
~\mathrm {(only~for~iron~K-lines)}
\end{eqnarray}
For iron K$_\alpha$ line, $u=2$, $l=1$. 
We emphasize again that the photon energy 
$\varepsilon_{lu}$ in Eqs. (21) is
in units erg (the CGSE units) rather than KeV (1 KeV=1.602$\times 10^{-9}$~ergs).
Eq.(20) gives a broad and
asymmetric profile of Cerenkov line with a small Cerenkov
line-redshift, as shown in Fig.3.

Inserting Eq.(20) into the integral
$I^c=\int_0^{y_{lim}}I_y^c{\mathrm d}y$, we obtain the total line intensity
$I^c$ in optically thick case
\begin{equation}
I^c=Y\left [\ln(1+X^2 )-2\left (1-\frac{\arctan X}{X}\right )\right ] \qquad ({\mathrm erg/sec.\cdot cm^{2} \cdot str.})
\end{equation}
where $Y\equiv\frac{N_e}{2}\frac{C_1}{k_{\mathrm {bf}}}\propto N_e$, $N_e$
is the density of relativistic electrons.
$X\equiv\sqrt\frac{k_{\mathrm {bf}}}{C_2}C_0\gamma_c^2\propto \gamma_{c}^{2}N_{\mathrm {Fe}}$, 
$N_{\mathrm {Fe}}$,
$\gamma_c$ represent the density of iron
ions in gas and the typical energy of relativistic electrons,
respectively. Therefore the total Cerenkov line intensity $I^c$ is
determined by the parameters $N_e$, $N_{\mathrm {Fe}}$, and $\gamma_c$.
Inserting the atomic parameters of iron ions into
Eq.(21), for the iron $K\alpha$
line, we obtain
\begin{eqnarray}
&&X=6.49\times10^{-28}g_2\sqrt{\frac{S_2}{g_2}\left (\frac{S_1}{g_1}-\frac{S_2}{g_2}\right )}
    N_{\mathrm {Fe}}\gamma_c^2 \nonumber \\
&&Y=0.16\times\frac{g_2}{S_2}\left (\frac{S_1}{g_1}-\frac{S_2}{g_2}\right )N_e
\end{eqnarray}

From eq. (23) we see, in a physically reasonable environment of AGNs, we usually have 
$X< 1$, or equivalently, $N_{\mathrm {Fe}}\gamma_c^2 <10^{27}$,Therefore, 
for optically thick and $X< 1$ case, Eq. (22) is simplified as
\begin{equation}
I_{K\alpha}^c\approx \frac{1}{3}YX^2
\end{equation}
$X$, $Y$ are given by Eq.(23) for iron K$_{\alpha}$ line. In another extremely case, $X>1$, or equivalently,
$N_{\mathrm {Fe}}\gamma_c^2>10^{27}$,
Eq.(22) is simplified as
\begin{equation}
I_{K\alpha}^c\approx2Y\left (\ln X-1\right )\approx 2Y \qquad 
\mathrm {for}~X>1 ~{\mathrm case}
\end{equation}

%%%%%%%%%%%%%%%%%%%%%%%%%%%%%%%%%%%%%%%%%%%%%%%%%%%%%%%%%%%%%%%%%%%%%%%%%%
\newpage
\begin{table}
\caption{Combinations of Iron Density, Characteristic Energy and Density of Relativistic
Electrons  for Calculation of  Cerenkov Luminosity of Iron K$_{\alpha}$ Line}
\begin{center}
         \begin{tabular}{|c|c|c|}
          \hline \hline
           \ $N_{\mathrm {Fe}} \mathrm (cm^{-3})$ & $\gamma_{c}$ & $N_{e} \mathrm (cm^{-3})$  \\
           \hline
           $10^{14}$ & $10^{6}$ & $10^{10}$ \\
           $10^{15}$ & $3\times 10^{5}$ & $10^{10}$ \\
           $10^{16}$ & $10^{5}$ & $10^{10}$ \\
           $10^{17}$ & $3\times 10^{4}$ & $10^{10}$ \\
           $10^{18}$ & $10^{4}$ & $10^{10}$ \\
           \hline \hline
         \end{tabular}
   \end{center}

\end{table}

\setcounter{figure}{0}

\newpage
\begin{figure}
\centerline{\psfig{figure=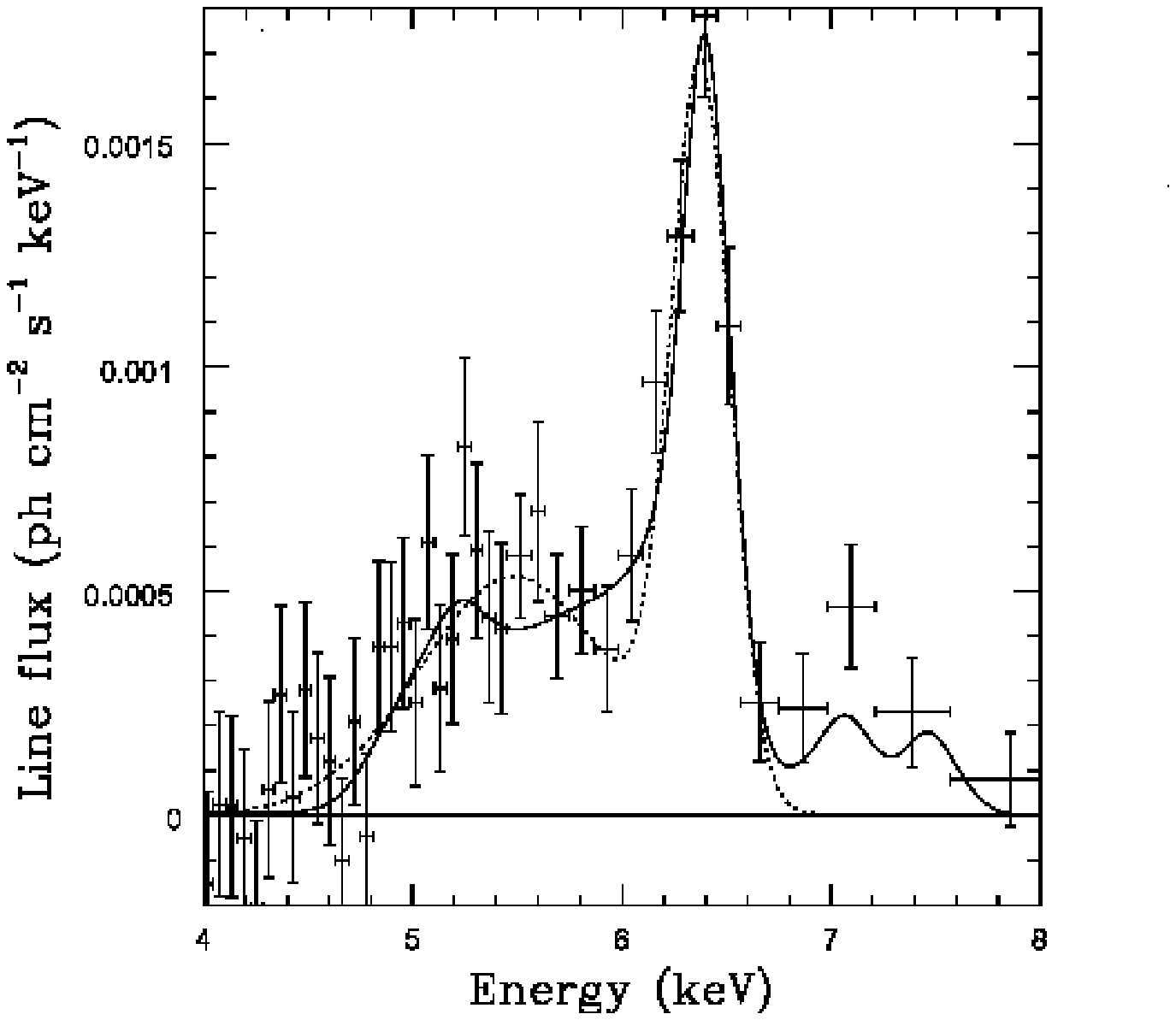,width=9.0cm,height=9.0cm}}
\centerline{\psfig{figure=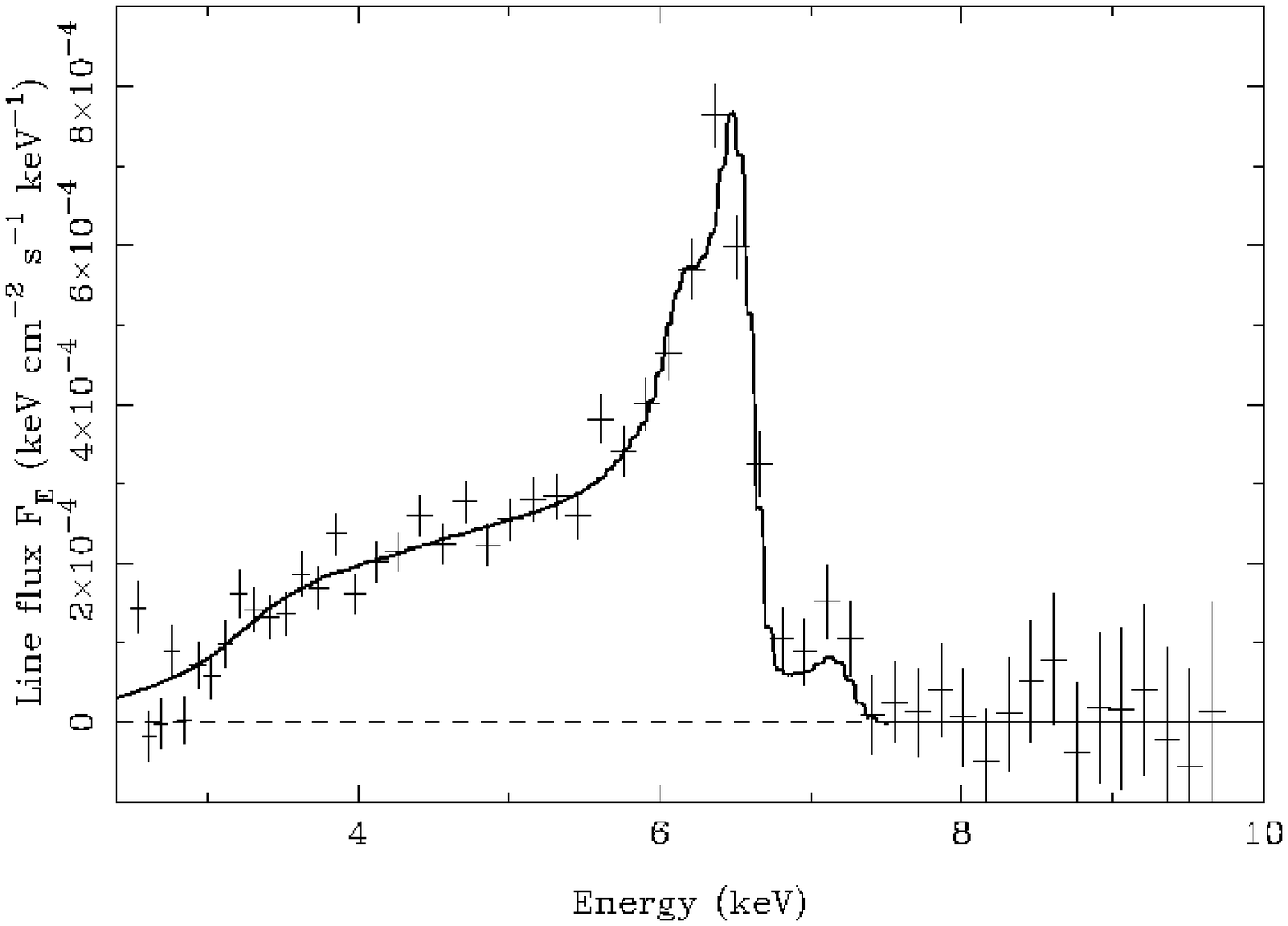,width=11.0cm,height=9.0cm}}
\caption{The observed iron K$_{\alpha}$ line of Seyfert 1 galaxy 
NGC 4151(upper panel,taken from Wang et al. 2001) and 
MCG-6-30-15(lower panel, taken from Fabian et al. 2002). 
The small line-like hump at $\sim$7-8KeV(in rest frame) is due to the 
iron $K_{\beta}$ line emission.}
 %\label{fig:frames}
\end{figure}

\newpage
\begin{figure}
\centerline{\psfig{figure=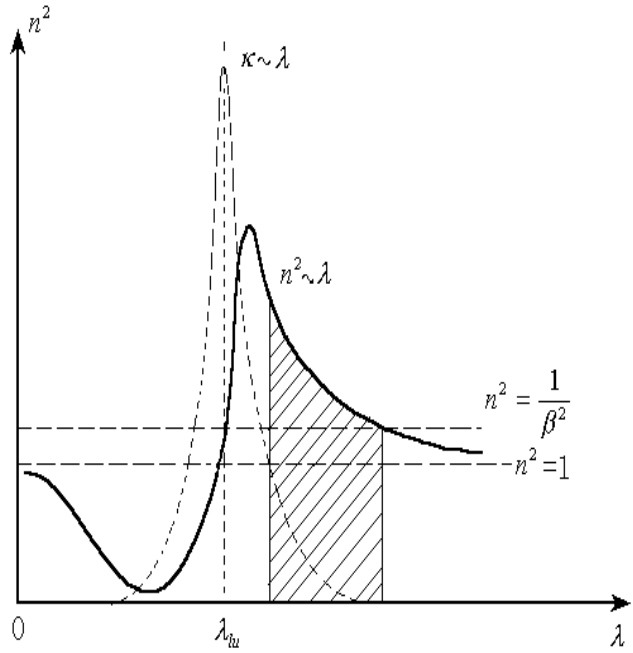,width=11.0cm,height=11.0cm}}
\caption{
Schematic sketch of the dispersion curve of gas $n^{2}\sim \lambda$, 
and the curve $\kappa \sim \lambda$, where $\kappa$ is the extinction 
coefficient of gas, which relates with the line-absorption coefficient 
by a simple formula $k_{\lambda}=\frac{4\pi}{\lambda}\kappa_{\lambda}$.
The Cerenlov radiation survives in the shaded narrow region
where the Cerenkov radiation
condition $n\geq 1/\beta$ is satisfied, and where the extinction is small.
The shaded region is narrow, making the emerging radiation appear more like a line
emission than a continuum.
 }
\end{figure}

\newpage
\begin{figure}
\centerline{\psfig{figure=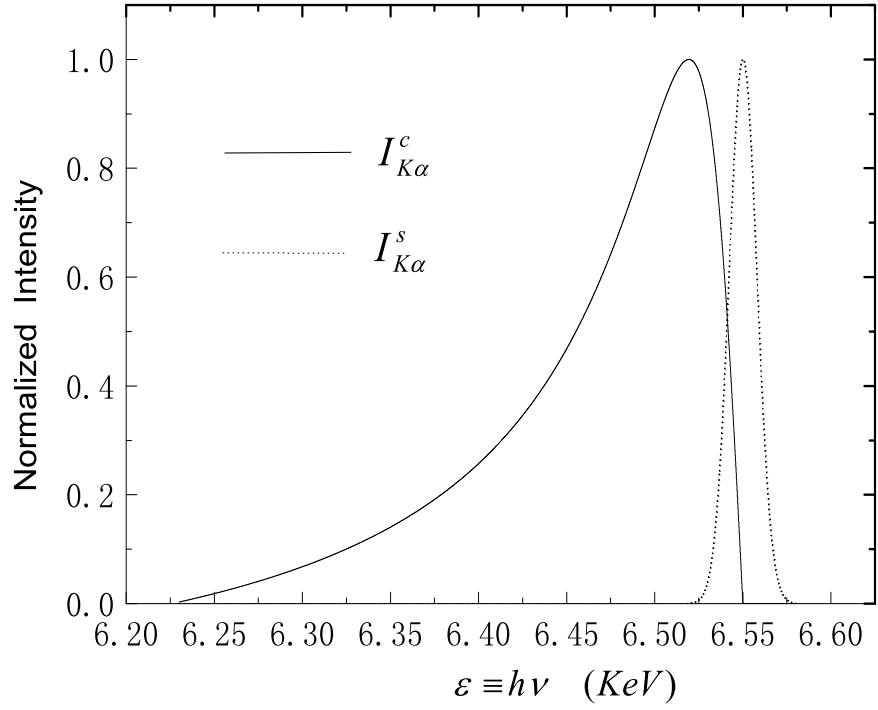,width=13.0cm,height=11.0cm}}
 \caption{
 The calculated profile of Cerenkov line $I_{K_{\alpha}}^{c}\sim \varepsilon$ of
 iron ion Fe$^{+21}$ in optically thick case, assuming $N_{\mathrm {Fe}}=10^{17} \mathrm{cm^{-3}}$ and
$\gamma_{c}=2\times 10^{5}$, where $\varepsilon \equiv h\nu$
is the energy of line photon. The Cerenkov line-profile is broad, asymmetric,
and redshifted. The profile of a normal line by spontaneous transition
$I_{K_{\alpha}}^{s}\sim \varepsilon$ is also plotted for comparison.
 }
\end{figure}

\newpage
\begin{figure}
\centerline{\psfig{figure=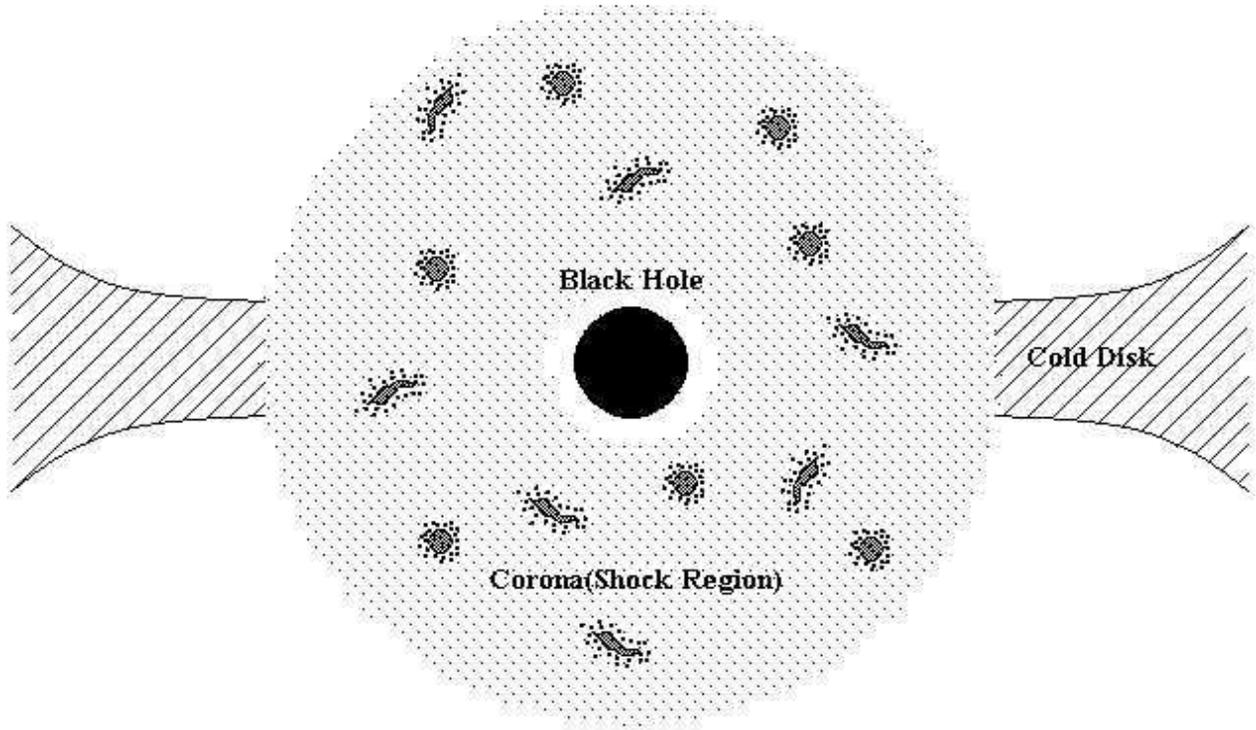,width=17.0cm,height=10.0cm}}
 \caption{
Schematic sketch of the emission region of iron K$_{\alpha}$ line around
 a central supermassive black hole of AGN.
The shaded circles or strips represent cloudlets or filaments 
consist of cold dense gas. The dotted region is the hot, 
rarified corona where the tiny dots and the black dots represent 
thermal electrons and relativistic electrons, respectively. 
The thermal electrons have an uniform distribution but the 
relativistic electrons have a high concentration around the clouds 
and/or filaments.}
\end{figure}

\end{document}